\documentstyle[amssymb,preprint,eqsecnum,aps,prb,amstex]{revtex}

\begin{document}
\title{Magnetic-history-dependent nanostructural and resistivity changes in Pr$%
_{0.5}$Ca$_{0.5}$Mn$_{0.98}$Cr$_{0.02}$O$_3$}
\author{M. Hervieu, R.Mahendiran, A. Maignan, C. Martin, and B. Raveau}
\address{Laboratoire CRISMAT, ISMRA, Universit\'{e} de Caen, 6 Boulevard du\\
Mar\'{e}chal Juin, 14050 Caen-Cedex, France.}
\maketitle

\begin{abstract}
We show that nanostructure and resistivity of Pr$_{0.5}$Ca$_{0.5}$Mn$_{0.98}$%
Cr$_{0.02}$O$_3$ are sensitive to whether the sample is zero-field-cooled
(ZFC) or field-cooled (FC) either in the `self-magnetic field (H $%
\thickapprox $ 2 T)' of the electron microscope or under the external
magnetic field of 2 T. FC resistivity at H = 2 T is lower than ZFC values
below 140 K. The average value of the charge-orbital modulation vector (q =
0.44) of the FC crystallites is lower than that of the ZFC crystallites (q =
0.48) and the FC crystallites exhibit numerous defects like
discommensuration, dislocations and regions with loss of superstructures
compared to the ZFC crystallites.

PACS NO: 61.14.-x, 75.30.Vn, 75.60.Ee\qquad
\end{abstract}

\smallskip

\newpage \quad \quad The renaissance of interest in charge ordered
manganites like Nd$_{0.5}$Sr$_{0.5}$MnO$_3$ is due to insulator-metal
transition driven by melting of charges under external magnetic fields\cite
{Rao}. Charge ordering (CO) manifests itself as regular ordering of Mn$^{3+}$%
:t$_{2g}^3$e$_g^1$ and Mn$^{4+}$: t$_{2g}^3$ ions and ordering of d$_z^2$
orbital at Mn$^{3+}$ sites. The low temperature phase of a perfect CO (Mn$%
^{3+}$/Mn$^{4+}$ = 1:1) system is a CE-type antiferromagnet, but some of the
manganites exhibit phase seperation into CO (antiferromagnetic, AF) and
non-CO (ferromagnetic, FM) regions\cite{Moreo}. Electron microscopy is one
of the powerful techniques to characterize the lattice distortion generated
by CO and phase separation\cite{Mori}. Neverthless, sample in the electron
microscope is under the influence of about 2 T (`self-magnetic field') and
CO in some of the manganites is sensitive to this `self-magnetic field' as
found recently in Pr$_{0.7}$Ca$_{0.3}$MnO$_3$\cite{Hervieu}. Coexistence of
ordered regions (correlated to the CO-AF phase) and non-ordered regions
(correlated to the FM phase) over wide x range ( 0.01 $\leq $ x $\leq $
0.05) has been reported\cite{Mahi,Kimura} in Pr$_{0.5}$Ca$_{0.5}$Mn$_{1-x}$Cr%
$_x$O$_3$. In this letter, we report unusual magnetic history dependence of
resistivity and nanostructural changes in Pr$_{0.5}$Ca$_{0.5}$Mn$_{0.98}$Cr$%
_{0.02}$O$_3$.

\smallskip

Resistivity ($\rho $) and magnetization ($M$) of polycrystalline Pr$_{0.5}$Ca%
$_{0.5}$Mn$_{0.98}$Cr$_{0.02}$O$_3$ were measured in two modes: In ZFC and
FC modes, data were taken while warming from 5 K in presence of a known
field (H) after cooling in zero-field and under H, respectively from 300 K.
In order to obtain the ZFC condition in electron microscopy study, the
objective lens of the microscope was cut off at 300 K before introducing the
sample and the temperature was decreased to 92 K. After one hour at 92 K,
the objective lens was switched on. Numerous crystallites have been
characterized using electron diffraction (ED) in order to determine the
average value of the charge modulation vector {\bf q} and lattice imaging
for the distribution of fringes. Different zones of different crystallites
have been located. Then the temperature was raised to 300 K keeping
objective lens switched on, again cooled to 92 K keeping exactly the same
current density in order to obtain the FC mode. Exactly the same areas of
the same crystallites were re-observed and the results are compared to those
recorded under ZFC condition.

\smallskip

The temperature dependence of M for H = 0.01 T, 2 T and $\rho $ for H = 0 T
and 2 T are shown in Figs. 1(a) and Fig. 1(b) respectively. We use solid
curve for the ZFC and dashed curve for the FC modes. The peak in M around
230 K in both H = 0.01 T and H = 2 T signals ordering of charges and
orbitals in the paramagnetic phase. A short range FM order is established
below T$_C$ = 130 K as suggested by the difference in ZFC and FC
magnetization at H = 0.01 T. $\rho $(0T) starts to decrease not at T$_C$ but
around 80 K, close to the temperature where the ZFC magnetization also
reaches a maximum. The decrease of $\rho $(T) below 80 K can be thus
considered due to the percolation of FM clusters whose moments are blocked
below this temperature. The peak in $\rho $(T) shifts to 125 K under H = 2 T
and the magnetoresistance ([$\rho _{ZFC}$(0 T)-$\rho _{FC}$(2 T)] /$\rho
_{FC}$(0T) ) at 100 K is as large as 73 \%. The surprising result is the
deviation of the ZFC and FC-$\rho $(T) below $\thickapprox $ 135 K and lower
values of the FC-$\rho $(T). The FC-$\rho $(T) exhibits a peak at T$_p$ ( =
128 K) which is higher than that of ZFC-$\rho $(T) (T$_p$ = 116 K). Under H
= 2 T, T$_C$ increases to 140 K, the maximum in ZFC-M(T) shifts down $%
\thickapprox $ 30 K and, FC-M(T) increases without saturation down to 5 K.
This suggests that long-range FM order is not reached even at H = 2 T and
canting of spins in the CO-AF phase contribute to non-saturation of M. We
find that irreversibility in $\rho $(T) and M(T) presists even at H = 7 T in
this compound but becomes negligibe with increasing Cr content \cite{Mahi2}
but, the details are beyond the scope of this letter.\quad

\smallskip

Since $\rho $(T) values are lower when it is field--cooled, we can expect
charge-ordering to be less stable in the FC state than in the ZFC state. The
periodic lattice distortion due charge-orbital ordering generates satellites
in Electron Diffraction (ED) patterns. The charge-orbital modulation vector
exhibit a value q = 0.5-$\varepsilon $ , where the incommensurability ($%
\varepsilon $) is zero for a perfect 1 : 1 charge ordering. Fig. 2(a) and
Fig. 2(b) respectively show the ED patterns at 92 K for the ZFC and FC
modes. The average q value of the FC crystallites is close to 0.44 which is
significantly lower than the average value q = 0.48 of the ZFC crystallites.
This is clearly observed in the [010] ED patterns through the amplitude of
the satellite splitting (see small black arrows). The satellites of the ZFC
crystallites are also more intense than the ones observed in the FC
crystallites. The corresponding lattice images are compared in Fig. 3(a) and
3(b). In the ZFC case (Fig. 3(a)), in addition to majority fringes of
periodicity 2a$_p\sqrt{2}$ = 10.8 \AA\ , we also find a few fringes of
higher periodicty with 3a$_p\sqrt{2}$= 16.2 \AA\ at irregular positions in
agreement with average q = 0.48. However, in the FC case (Fig. 3(b),
numerous defects are observed in the arrangements of fringes:
discommensurations ( for example, area noted D) which are easily seen by
viewing the images at grazing incidence and local losses of superstructure
(for example, area within open arrows). To our best of knowledge, this is
the first time influence of magnetic field on nanostructure has been
reported. It is important to recall that artefacts due to bad stabilization
of the temperature or hysteresis effect can generate the difference in the q
values observed for exactly the same area. We repeated our experiments
several times and all our results are identical. The satellites are stable
under electron beam with long exposure contrary to what was found\cite
{Hervieu} in Pr$_{0.7}$Ca$_{0.3}$MnO$_3$. This suggests that CO is more
stable in the present compound.

\smallskip

\smallskip The above results unequivocally suggest that magnetic history of
the sample affects not only the resistivity but also nanostructure. Why is
it so ?. Recently, we have suggested\cite{Mahi} that Cr$^{3+}$ :t$_{2g}^3$
ions randomly occupy Mn$^{3+}$:t$_{2g}^3$e$_g^1$ sites with their t$_{2g}^3$
spins opposite to the replaced Mn$^{3+}$ ions. The random substitution of Cr$%
^{3+}$ ions brings about e$_g$-orbital deficiency at the original Mn$^{3+}$%
-sites and affects cooperative Jahn-Teller distortion of Mn$^{3+}$O$_6$
octahedras. Hence, Cr$^{3+}$ ions act as random impurities for spin as well
as lattice degrees of freedom. Antiferromagnets with random impurities
(`diluted Ising antiferromagnets') show long range AF order when they are
zero field cooled but break into domains upon field cooling\cite{Imry}.
While the sizes of the domains in theoretical models \cite{Imry} are purely
determined between volume and magnetic surface energies, spin-charge-lattice
coupling\cite{Ibarra} also plays important role in manganites. Hence, the
long range d$_{z^2-}$ orbital ordering is also broken by the formation of AF
spin domains which also causes orbital domains to form. Such orbital domains
lead to discommensuration and dislocations observed experimentally. Field
cooling is also found to enhance the size of FM domains \cite{Kimura,Mahi2}%
which are clearly visible in lattice images as regions without
superstructure. The breaking of CO-AF matrix into domains and growth of FM
domains lead to low resistivity under field cooling.

\smallskip

\smallskip In conclusion, we have shown that nanostructure and resistivity
of \quad Pr$_{0.5}$Ca$_{0.5}$Mn$_{0.98}$ \quad \quad Cr$_{0.02}$O$_3$ are
strongly magnetic field history dependent. Our study clearly suggests that
electron microscopy can be used to tune nanostructure in phase separated
manganites.

\smallskip

Acknowledgements:

\smallskip

R. M thanks MNENRT (France) for financial support and acknowledges Prof. T.
V. Ramakrishnan for his encouragement.

\begin{center}
\smallskip

FIGURE CAPTIONS
\end{center}

\begin{description}
\item[Fig.1]  : (a) Temperature dependence of magnetization at H = 0.01 T
and 2 T in ZFC (solid curve) and FC (dashed curve). (b). $\rho $(T) under
ZFC (solid curve) and FC (dashed curve) for H = 2 T and in absence of field
(H = 0 T) modes. T$_{CO}$: Charge ordering temperature, T$_C$: ferromagnetic
Curie temperature.

\item[Fig.2]  : [010] ED patterns at 92 K underZFC (a) and FC (b) modes.
Note that satellites marked by double arrows are clearly splitted and less
intense in FC mode than in ZFC mode.

\item[Fig.3]  : [010] lattice images at 92 K under ZFC (a) and FC (b) modes.
Regions with loss of superstructures are denoted by open arrows and one
region with discommensuration is marked by D.
\end{description}

\end{document}